\documentclass[singlecolumn,epjc3]{svjour3}
\smartqed  
\RequirePackage{graphicx}
\usepackage{amsmath,amssymb}
\RequirePackage{mathptmx}
\journalname{Eur. Phys. J. C}

\begin{document}

\title{Study on anisotropic strange stars in $f(\mathbb{T},\mathcal{T})$ gravity}

\author{I.G. Salako\thanksref{e1,addr1} \and M. Khlopov\thanksref{e2,addr2} \and Saibal Ray\thanksref{e3,addr3} \and M.Z. Arouko\thanksref{e4,addr4} \and Pameli Saha\thanksref{e5,addr5} \and Ujjal Debnath\thanksref{e6,addr6}.}

\thankstext{e1}{e-mail: inessalako@gmail.com}
\thankstext{e2}{e-mail: khlopov@apc.in2p3.fr}
\thankstext{e3}{e-mail: saibal@iucaa.ernet.in}
\thankstext{e4}{e-mail: maximearouko55@gmail.com}
\thankstext{e5}{e-mail: pameli.saha15@gmail.com}
\thankstext{e6}{e-mail: ujjaldebnath@gmail.com}

\institute{D\'epartement de Physique, Universit\'e Nationale d’Agriculture, 01 BP 55 Porto-Novo, Benin; Institut de Math\'ematiques et de Sciences Physiques (IMSP), Universit\'e  d'Abomey-Calavi Porto-Novo, 01 BP 613 Porto-Novo, Benin\label{addr1} \and
National Research Nuclear University, MEPHI (Moscow Engineering Physics Institute), Moscow 115409, Russia; Universit\'e de Paris, CNRS,
Astroparticule et Cosmologie, F-75013 Paris, France and Institute of Physics, Southern Federal University, Rostov on Don 344090, Russia\label{addr2} \and  Department of Physics, Government College of Engineering \& Ceramic Technology, Kolkata 700010, West Bengal,
India\label{addr3} \and Département de Physique, Université d’Abomey-Calavi, BP 526 Calavi, Benin\label{addr4} \and Department of
Mathematics, Indian Institute of Engineering Science and Technology, Shibpur, Howrah 711103, West Bengal, India\label{addr5} \and Department of Mathematics, Indian Institute of Engineering Science and Technology, Shibpur, Howrah 711103, West Bengal, India\label{addr6}}

\date{Received: date / Accepted: date}

\maketitle

\begin{abstract}
In this work, we study the existence of strange star in the background of $f(\mathbb{T},\mathcal{T})$ gravity in the Einstein spacetime geometry, where $\mathbb{T}$ is the torsion tensor and $\mathcal{T}$ is the trace of the energy-momentum tensor. The equations of motion are derived for anisotropic pressure within the spherically symmetric strange star. We explore the physical features like energy conditions, mass-radius relations, modified TOV equations, principal of causality, adiabatic index, redshift and stability analysis of our model. These features are realistic and appealing to further investigation of properties of compact objects in $f(\mathbb{T},\mathcal{T})$ gravity as well as their observational signatures.
\end{abstract}

\keywords{general relativity; alternative gravity; compact stars; anisotropy}

\section{Introduction}
In relativistic astrophysics, compact stars are more compact (i.e.,  possess larger mass and smaller radius) than ordinary stars, has received much attention to the researchers to study their ages, structures and evolutions. Neutron star is the final stage of a gravitationally collapsed star which, after exhausting all its thermo-nuclear fuel, gets stabilized by degenerate neutron pressure. A massive neutron star may be again collapsed into black hole but lower mass neutron star may be converted into quark star. The strange quark matter~\cite{Bod,Tera,Wit} consists of $u$, $d$, $s$ quarks and some electrons, which may be more stable than the ordinary nuclear matter. Actually, neutron star consists of neutrons~\cite{BWZF34} but strange star consists of quarks or strange matters.

In 1916, Schwarzschild et al.~\cite{1SK16,2SK16} have first given the interior stellar solution for isotropic fluid (having equal radial pressure ($p_{r}$) and transversal pressure ($p_{t}$)). But much later in 1972, Ruderman~\cite{RR72} first observed that the interior geometry of the nuclear matter of steller system with density of order $10^{15} gm/cc$ possesses anisotropic behavior (i.e., $p_{r}\neq p_{t}$). After that Herrera et al.~\cite{HLSNO97} have analyzed the local anisotropic nature for self gravitating systems. It is interesting to note that in the experimental area Hewish et al.~\cite{HABSJPJDH68} explained the observed compactness of many astrophysical objects (rotating neutron stars) like Her X-$1$, $4$U $1820-30$, RXJ $1856-37$ and SAX J $1808.4-3658$. Thus the theoretical modeling gradually became complimentary to the theoretical explanations.

The study of anisotropic stars in the background of General Relativity (GR) is important in the current research. Hossein et al.~\cite{HSKM12} have studied the stable structure of stellar objects for anisotropic system. In the context of the now Standard cosmology, leading beyond the standard model of particle theory, such analysis inevitably involves additional elements in the standard model of gravity. Consequently, Kalam et al.~\cite{KMRFHSKM14} have studied the anisotropic neutron star model by considering quintessence dark energy. To study compact star models, the researchers have used well known Krori-Barua (KB) metric~\cite{Krori,Kalam1}. Paul et al.~\cite{PBCDR14} have analyzed a new exact solution for compact star. Several astrophysical natures of quark star and neutron star have been investigated in refs.~\cite{WE84,CKSDZGLT98}. Bhar~\cite{BP14} has proposed a stable anisotropic quintessence strange star model whereas Abbas et al.~\cite{Abbas9} have studied anisotropic strange star for quintessence dark energy model. Using MIT bag model, Rahaman et al.~\cite{RFCKKPKF14} have observed the existence of strange star. Also, by considering MIT bag model, the stability of strange star with anisotropic fluid has been analyzed by Arba$\tilde{n}$il and Malheiro~\cite{AJDV16}. Furthermore, for MIT bag model, Murad~\cite{MMH16} has studied the anisotropic charged strange star.

The study of anisotropic stars in the background of modified gravity theory is of great interest to many researchers~\cite{Mak,Abbas4,Abbas2,Abbas5,Abbas3,Abbas1}. The compact stars in $f(R)$, $f(G)$, $f(\mathbb{T})$ and $f(R,G)$ gravity theories have been studied by several authors~\cite{Abbas10,Abbas7,Abbas6,AG15,Deb2018,Das2020,Nojiri2005}, where $R$, $G$ and $\mathbb{T}$ are Ricci scalar, Gauss-Bonnet scalar and the torsion scalar, respectively. Strange stars in $f(\mathbb{T})$ gravity with MIT bag model have been studied in~\cite{A1}. Compact stars, in the form of neutron star as well as quark star, in Rastall theory of gravity model have been studied by several researches~\cite{R1,R2,R3,R4,R5,R6} whereas strange stars with MIT bag model in the Rastall gravity have been studied in Ref.~\cite{A2}. Charged anisotropic collapsing stars with heat flux in $f(R)$ gravity has been investigated by Nazar and Abbas~\cite{A3}.

Studies on anisotropic stars in different types of modified gravity theories are available in literature~\cite{S1,M1,M2,M3,P1,AS1}. Saha and Debnath~\cite{SPDU18} have investigated anisotropic stars in $f(\mathbb{T})$ gravity with modified Chaplygin gas. The anisotropic compact stars with quintessence field and modified Chaplygin gas in the framework of $f(\mathbb{T})$ gravity model has been studied by Saha and Debnath~\cite{De}. On the other hand, strange stars as well as gravastars in $f(R,\mathcal{T})$ gravity have been studied in~\cite{D1,D2,SB10,Moraes2016,Das2017,Deb2019a,Deb2019b,Deb2019c,Biswas2020}, where $\mathcal{T}$ is the trace of the energy-momentum tensor. Pace et al.~\cite{Pace1,Pace2} have studied quark star and neutron star models in $f(\mathbb{T},\mathcal{T})$ gravity theory whereas gravastars has been investigated by Ghosh et al. in the same alternative theory of gravity~\cite{Ghosh2020}.

Thus our motivation to study anisotropic strange stars in $f(\mathbb{T},\mathcal{T})$ gravity lies in the above background in the presence of
Einstein spacetime within the spherically symmetric geometry. The organization of the paper is as follows: In Sect. 2, the mathematical formulation of $f(\mathbb{T},\mathcal{T})$ gravity in Einstein spacetime is given. In Sect. 3, we investigate the basic stellar equations in the framework of $f(\mathbb{T},\mathcal{T})$ gravity. In Sect. 4, we find out the solutions to the Einstein field equations due to anisotropic fluid source. In Sect. 5, we explore the physical features like energy conditions, mass-radius relations, modified TOV equations, principal of causality, adiabatic index, redshift and stability analysis of our model. In Sect. 6, we deliver the conclusions of the work.

\section{Basic mathematical formalism of $f(\mathbb{T},\mathcal{T})$ gravity and the Einstein field equations}
The modified theories of teleparallel gravity  are those for which the scalar torsion of teleparallel action is substituted by an arbitrary function of this latter. As it is done in teleparallel, the modified versions of  this theory are also described by the orthonormal tetrads and it's components are defined on the tangent space of each point of the manifold. The line element is written as
\begin{equation}
ds^2=g_{\mu\nu}dx^\mu dx^\nu=\eta_{ij}\theta^i\theta^j,\label{eq2}
\end{equation}
with the following definitions
\begin{equation}
dx^{\mu}=e_{i}^{\;\;\mu}\theta^{i}; \,\quad \theta^{i}=e^{i}_{\;\;\mu}dx^{\mu}.\label{eq3}
\end{equation}

Note that  $\eta_{ij}=diag(1,-1,-1,-1)$ is the Minkowskian metric and the $\{e^{i}_{\;\mu}\}$ are the components of the tetrad which
satisfy the following identity:
\begin{eqnarray}
e^{\;\;\mu}_{i}e^{i}_{\;\;\nu}=\delta^{\mu}_{\nu},\quad e^{\;\;i}_{\mu}e^{\mu}_{\;\;j}=\delta^{i}_{j}.\label{eq4}
\end{eqnarray}

In General Relativity, one use the following Levi-Civita's connection which  preserves the curvature whereas the torsion vanishes, such as
\begin{equation}
\overset{\circ}{\Gamma }{}_{\;\;\mu \nu }^{\rho } =
\frac{1}{2}g^{\rho \sigma }\left(
\partial _{\nu} g_{\sigma \mu}+\partial _{\mu}g_{\sigma \nu}-\partial _{\sigma}g_{\mu \nu}\right).\label{eq5}
\end{equation}

But in the teleparallel theory and its modified version, one keeps the scalar torsion by using Weizenbock's connection
defined as:
\begin{eqnarray}\label{eq6}
\Gamma^{\lambda}_{\mu\nu}=e^{\;\;\lambda}_{i}\partial_{\mu}e^{i}_{\;\;\nu}=-e^{i}_{\;\;\mu}\partial_\nu e_{i}^{\;\;\lambda}.
\end{eqnarray}

From this connection, one obtains the geometric objects. The first is the torsion as defined by
\begin{equation}
T^{\lambda}_{\;\;\;\mu\nu}= \Gamma^{\lambda}_{\mu\nu}-\Gamma^{\lambda}_{\nu\mu},\label{eq7}
\end{equation}
from which we define the contorsion as
\begin{eqnarray}
K_{\;\;\mu \nu }^{\lambda} \equiv \Gamma _{\;\mu \nu }^{\lambda }
-\overset{\circ}{\Gamma }{}_{\;\mu \nu }^{\lambda}=\frac{1}{2}(T_{\mu }{}^{\lambda}{}_{\nu }
+ T_{\nu}{}^{\lambda }{}_{\mu }-T_{\;\;\mu \nu }^{\lambda}),\label{eq8}
\end{eqnarray}
where the expression $\overset{\circ }{\Gamma }{}_{\;\;\mu \nu }^{\lambda}$ designs the above defined connection. Then we can write
\begin{equation}
K^{\mu\nu}_{\;\;\;\;\lambda}=-\frac{1}{2}\left(T^{\mu\nu}_{\;\;\;\lambda}-T^{\nu\mu}_{\;\;\;\;\lambda}+T^{\;\;\;\nu\mu}_{\lambda}\right).\label{eq9}
\end{equation}

The two previous geometric objects (the torsion and the contorsion) are used to define another tensor by
\begin{equation}
S_{\lambda}^{\;\;\mu\nu}=\frac{1}{2}\left(K^{\mu\nu}_{\;\;\;\;\lambda}+
\delta^{\mu}_{\lambda}T^{\alpha\nu}_{\;\;\;\;\alpha}-\delta^{\nu}_{\lambda}T^{\alpha\mu}_{\;\;\;\;\alpha}\right).\label{eq10}
\end{equation}

The torsion scalar is usually constructed from torsion and contorsion as follows:
\begin{equation}
\mathbb{T} = S_\sigma^{~\mu\nu} T^\sigma_{~\mu\nu}.\label{eq11}
\end{equation}

In the modified versions of teleparallel gravity, one can use a general algebraic function of scalar torsion instead of the scalar torsion only as it is done in the initial theory. So, the modified action can be given by
\begin{equation}
\mathbb{S}= \int d^{4}x~~e \left[\frac{\mathbb{T}+f(\mathbb{T},\mathcal{T})}{16\pi}+\mathcal{L}_{m}\right],\label{3.1}
\end{equation}
where $T_{\mu\nu}$ is the energy-momentum tensor of the strange quark matter (SQM) distribution and  ${\mathcal{L}}_m$ represents the Lagrangian for the matter distribution.

We define $T_{\mu\nu}$ as follows
\begin{eqnarray}\label{3.2}
T_{\mu\nu}=-\frac{2}{\sqrt{-g}}\frac{\delta(\sqrt{-g} {{\mathcal{L}}_m})}{\delta g^{\mu\nu}},
\end{eqnarray}
and also define the trace of $T_{\mu\nu}$ as $\mathcal{T}=g^{\mu\nu}T_{\mu\nu}$. As ${\mathcal{L}}_m$ depends only on the metric tensor components $g_{\mu\nu}$ and not on their derivatives, so we find
\begin{eqnarray}\label{3.3}
T_{\mu\nu}=g_{\mu\nu} {\mathcal{L}}_m - 2 \frac{\partial {\mathcal{L}}_m}{\partial {g^{\mu\nu}}}.
\end{eqnarray}

Varying the action with respect to the tetrad, one obtains the equations of motion~\cite{Harko2014} as
\begin{eqnarray}\label{eq13}
&&[\partial_\xi(ee^\rho_a S^{\;\;\sigma\xi}_\rho)-ee^\lambda_a S^{\rho\xi\sigma} T_{\rho\xi\lambda}](1+f_{\mathbb{T}})
+ e e^\rho_a(\partial_\xi\mathbb{T})S^{\;\;\sigma\xi}_\rho f_{\mathbb{T}\mathbb{T}}\nonumber\\
&& +\frac{1}{4} e e^\sigma_a (\mathbb{T}) =- \frac{1}{4} e e^\sigma_a f -e e^\rho_a(\partial_\xi \mathcal{T})S^{\;\;\sigma\xi}_\rho f_{\mathbb{T}\mathcal{T}} \nonumber\\
&& +f_{\mathcal{T}}\;\left(\frac{e\,T^\sigma_{\;\;a}  + e e^\sigma_a \;p }{2}\right) + 4\pi\,e\,T^\sigma_{\;\;a}\;,
\end{eqnarray}
with
$f_{\mathcal{T}} = \partial f/\partial \mathcal{T} $,
$f_{\mathbb{T}} = \partial f/\partial\mathbb{T}$,
$ f_{\mathbb{T}\mathcal{T}} = \partial^{2}f/\partial\mathbb{T}\partial \mathcal{T}$,
$f_{\mathbb{T}\mathbb{T}}  = \partial^{2}f/\partial \mathbb{T}^{2}$
and $T^\sigma_{\;\;a}$ represents the stress-energy tensor for the anisotropic fluid distribution defined as
\begin{eqnarray}\label{3.6}
T_{\mu\nu}=\left(p_r - p_t \right)u_{\mu} u_{\nu} -p_t g_{\mu\nu} + \left( \rho+p_t \right)u_{\mu} u_{\nu},
\end{eqnarray}
where where $p_r$ and $p_t$ represent the radial and tangential pressures of the SQM distribution, $u_{\mu}$ is the four velocity which satisfies the conditions $u_{\mu}u^{\mu}=1$ and $u^{\mu} {\nabla}_{\nu} u_{\mu}=0$.

By using some transformations, we can establish the following relations:
\begin{eqnarray}\label{eq15}
e^a_\nu e^{-1}\partial_\xi(ee^\rho_a
S^{\;\;\sigma\xi}_\rho)-S^{\rho\xi\sigma}T_{\rho\xi\nu} = -\nabla^\xi S_{\nu\xi}^{\;\;\;\;\sigma}-S^{\xi\rho\sigma}K_{\rho\xi\nu},
\end{eqnarray}

\begin{eqnarray}\label{eq16}
G_{\mu\nu}-\frac{1}{2}\,g_{\mu\nu}\,\mathbb{T}
=-\nabla^\rho S_{\nu\rho\mu}-S^{\sigma\rho}_{\;\;\;\;\mu}K_{\rho\sigma\nu}.
\end{eqnarray}

Hence from the combination of Eqs.~(\ref{eq15}) and (\ref{eq16}), the field equation (\ref{eq13}) can be written as:
\begin{eqnarray}
\frac{1}{2}(1+f_{\mathbb{T}})\,G_{\mu\nu}=\frac{1}{4}g_{\mu\nu}\mathbb{T}\,(1+f_{\mathbb{T}})
-S_{\nu\mu}^{\;\;\;\;\lambda}\left(f_{\mathbb{T}\mathbb{T}}\partial_{\lambda}\mathbb{T}+f_{\mathbb{T}\mathcal{T}}\partial_{\lambda}\mathcal{T}\right)-\frac{1}{4}g_{\mu\nu}\left(f+\mathbb{T}\right) \nonumber \\ +\frac{1}{2}f_{\mathcal{T}}\left(T_{\nu\mu}+g_{\mu\nu}p_{t}\right) +4\pi G T_{\nu\mu},\label{eq20}
\end{eqnarray}

\begin{equation}
G_{\mu\nu}=8\pi G \;T_{\mu\nu}^{eff},\label{eq20}
\end{equation}
where
\begin{eqnarray}
T_{\mu\nu}^{eff}=\frac{1}{(1+f_{\mathbb{T}})} \Bigg \{ \frac{g_{\mu\nu}\mathbb{T}}{16\pi G}\,(1+f_{\mathbb{T}})
- \frac{S_{\nu\mu}^{\;\;\;\;\lambda}}{4\pi G} \left(f_{\mathbb{T}\mathbb{T}}\partial_{\lambda}\mathbb{T}+f_{\mathbb{T}\mathcal{T}}\partial_{\lambda}\mathcal{T}\right)-\frac{g_{\mu\nu}}{16\pi G}\left(f+\mathbb{T}\right) \nonumber\\+\frac{f_{\mathcal{T}}}{8\pi G} \left(T_{\nu\mu}+g_{\mu\nu}p_{t}\right)+ T_{\nu\mu} \Bigg \}.
\end{eqnarray}

The covariant derivative of Eq. (\ref{eq20}) reads as
\begin{eqnarray}
&&\nabla_{\mu} T_{\nu}^{\;\;\mu}=\frac{1}{\left(4\pi G+(1/2)f_{\mathcal{T}}\right)}\Bigg\{\left(f_{\mathbb{T}\mathbb{T}}\partial_{\sigma}\mathbb{T}+f_{\mathbb{T}\mathcal{T}}\partial_{\sigma}\mathcal{T}\right)e^{a}_{\;\;\nu}[e^{-1}\partial_{\lambda}(ee_{a}^{\;\;\alpha}S_{\alpha}^{\;\;\sigma\lambda})-e_{a}^{\;\;\alpha}T^{\lambda}_{\;\;\gamma\alpha}S_{\lambda}^{\;\;\gamma\sigma}]-\frac{1}{4}(1+f_{\mathbb{T}})\partial_{\nu}\mathbb{T}\nonumber\\
&&+\nabla_{\mu}S_{\nu}^{\;\;\mu\lambda}\left(f_{\mathbb{T}\mathbb{T}}\partial_{\lambda}\mathbb{T}+f_{\mathbb{T}\mathcal{T}}\partial_{\lambda}\mathcal{T}\right)+S_{\nu}^{\;\;\mu\lambda}\Big(f_{\mathbb{T}\mathbb{T}\mathbb{T}}\partial_{\mu}\mathbb{T}\partial_{\lambda}\mathbb{T}+f_{\mathbb{T}\mathbb{T}\mathcal{T}}\partial_{\mu}\mathcal{T}\partial_{\lambda}\mathbb{T}+f_{\mathbb{T}\mathbb{T}}\nabla_{\mu}\partial_{\lambda}\mathbb{T}+f_{\mathbb{T}\mathcal{T}\mathbb{T}}\partial_{\mu}\mathbb{T}\partial_{\lambda}\mathcal{T}\nonumber\\
&&+f_{\mathbb{T}\mathcal{T}\mathcal{T}}\partial_{\mu}\mathcal{T}\partial_{\lambda}\mathcal{T}+f_{\mathbb{T}\mathcal{T}}\nabla_{\mu}\partial_{\lambda}\mathcal{T}\Big)+\frac{1}{4}\left(f_{\mathbb{T}}\partial_{\nu}\mathbb{T}+f_{\mathcal{T}}\partial_{\nu}\mathcal{T}+\partial_{\nu}\mathbb{T}\right)-\frac{1}{2}\left(f_{\mathcal{T}\mathbb{T}}\partial_{\mu}\mathbb{T}+f_{\mathcal{T}\mathcal{T}}\partial_{\mu}\mathcal{T}\right)\left(T_{\nu}^{\;\;\mu}+\delta^{\mu}_{\nu}p_{t}\right)\nonumber\\
&&-\frac{1}{2}f_{\mathcal{T}}\partial_{\nu}p_{t}\Bigg\}. \label{eqnonconserv}
\end{eqnarray}

In the current work, we are focused on the existence of anisotropic strange stars in the extended teleparallel gravity and for this purpose we use a algebraic function according to observational data~\cite{Harko2014} $f\left(\mathbb{T},\mathcal{T}\right)=\varpi\, \mathbb{T}^n\,
\mathcal{T} -2  \Lambda$, where $\varpi, n$ and $\Lambda $ are arbitrary constants.

For the above mentioned functional form of $f(\mathbb{T},\mathcal{T})$ the Eq. (\ref{eq20}) reduces to
\begin{equation}
\,G_{\mu\nu}=8\pi  \;T_{\mu\nu}^{eff},\label{eq201}
\end{equation}
where
\begin{eqnarray}
T_{\mu\nu}^{eff}= g_{\mu\nu} \Bigg[\frac{\Big(-\varpi\,(\rho -p_r-2 p_t) +2  \Lambda \Big)}{16\pi }
+\frac{\varpi p_t}{8\pi }  \Bigg]+ T_{\nu\mu} \Bigg(1+ \frac{\varpi}{8\pi } \Bigg)
\end{eqnarray}

and

\begin{eqnarray}
\nabla_{\mu} T_{\nu}^{\;\;\mu}= \frac{1}{\Big(4\pi +(1/2)\varpi\Big)}\Bigg\{\frac{\varpi}{4}\left(\partial_{\nu}\mathcal{T}\right)-\frac{\varpi}{2} \partial_{\nu}p_{t}\Bigg\}\label{eqnonconserv}.
\end{eqnarray}

\section{Explicit stellar equations in $f(\mathbb{T}, \mathcal{T})$ gravity under the Einsteinian spacetime} \label{sec4}
To describe interior spacetime of the spherically symmetric static stellar system, we take metric as follows:
\begin{eqnarray}\label{3.13}
ds^2=e^{\nu(r)}dt^2-e^{\lambda(r)}dr^2-r^2(d\theta^2+\sin^2\theta d\phi^2),
\end{eqnarray}
where the metric potentials $\nu$ and $\lambda$ are the functions of the radial coordinate $r$ only.

In order to re-write the line element Eq. (\ref{3.13}) into the invariant form under the Lorentz transformations as in Eq. (\ref{eq2}),
we define the tetrad matrix $[e^{i}_{\;\mu}]$ as
\begin{equation}
\left[e^{i}_{\;\;\mu}\right]= diag \left[e^{\nu(r)/2},e^{\lambda(r)/2},r,r\sin\theta\right].\label{eq26}
\end{equation}

From  Eq. (\ref{eq26}), one can obtain $e=\det{\left[e^{i}_{\;\;\mu}\right]}=e^{(\nu+\lambda)/2}r^2 \sin\theta$ and using Eq. (\ref{eq11}) the  torsion scalar can be written as
\begin{equation}
\mathbb{T}(r)= \frac{2e^{-\lambda}}{r}\left(\nu^{\prime}+\frac{1}{r}\right).\label{eq27}
\end{equation}

Now the nonzero components of the Einstein tensors can be provided as follows:
\begin{equation}
G_0^{0}=\frac{e^{-\lambda}}{r^{2}}(-1+e^{\lambda}+\lambda'r),\label{eq28}
\end{equation}

\begin{equation}
G_1^{1}=\frac{e^{-\lambda}}{r^{2}}(-1+e^{\lambda}-\nu'r),\label{eq29}
\end{equation}

\begin{equation}
G_2^{2}=G_3^{3}=\frac{e^{-\lambda}}{4r}[2(\lambda'-\nu')-(2\nu''+\nu'^{2}-\nu'\lambda')r],\label{eq30}
\end{equation}
where primes stand for derivative with respect to the radial coordinate $r$ only.

Substituting Eq. \eqref{3.6} into Eq. \eqref{eq201}, we find the explicit form of the Einstein field equations for the interior
metric \eqref{3.13} as follows:
\begin{eqnarray}\label{3.21}
& \hspace{-1cm} {{\rm e}^{-\lambda }} \left( {\frac {\lambda^{{\prime}}}{r}}-\frac{1}{r^2}\right) +\frac{1}{r^2}= 8\pi \Bigg\{  \Bigg[\frac{\Big(-\varpi\,(\rho -p_r-2 p_t) +2  \Lambda \Big)}{16\pi }
+\frac{\varpi p_t}{8\pi }  \Bigg]+ \rho\Bigg(1+ \frac{\varpi}{8\pi } \Bigg)  \Bigg\}  =8\pi \rho^{\textit{eff}}, \\ \label{3.22}
& \hspace{-1cm} {{\rm e}^{-\lambda}} \left( {\frac {\nu^{{\prime}}}{r}}+\frac{1}{r^2}\right) -\frac{1}{r^2}=-8\pi \Bigg\{\Bigg[\frac{\Big(-\varpi\,(\rho -p_r-2 p_t) +2  \Lambda \Big)}{16\pi }
+\frac{\varpi p_t}{8\pi }  \Bigg]-p_r \Bigg(1+ \frac{\varpi}{8\pi } \Bigg)  \Bigg\}=8\pi  p_r^{\textit{eff}},
\\ \label{3.22a}
& \hspace{-0.1cm}
\frac{e^{-\lambda}}{4r}[2(\lambda'-\nu')-(2\nu''+\nu'^{2}-\nu'\lambda')r]
=-8\pi \Bigg\{\Bigg[\frac{\Big(-\varpi\,(\rho -p_r-2 p_t) +2  \Lambda \Big)}{16\pi }
+\frac{\varpi p_t}{8\pi }  \Bigg]-p_t \Bigg(1+ \frac{\varpi}{8\pi } \Bigg)  \Bigg\} \nonumber\\ =8\pi  p_t^{\textit{eff}}.
\end{eqnarray}

Here $\rho^{\textit{eff}}$, $p_r^{\textit{eff}}$ and $p_t^{\textit{eff}}$ represent the effective energy density, radial pressure and tangential pressure for our system and given as
\begin{eqnarray}\label{3.23}
& \rho^{\textit{eff}}=\rho + \frac{\varpi}{16\pi} (\rho + p_r + 4 p_t + 2 \Lambda),
\\ \label{3.23a}
& p_r^{\textit{eff}}=p_r+ \frac{\varpi}{16\pi} (\rho + p_r - 4 p_t - 2 \Lambda),
\\ \label{3.23aa}
& p_t^{\textit{eff}}=p_t + \frac{\varpi}{16\pi} (\rho - p_r + 2 p_t - 2 \Lambda).
\end{eqnarray}

We assume that the SQM distribution inside the strange stars is governed by the simple phenomenological MIT Bag model equation of state
(EOS)~\cite{Chodos1974}. In the Bag model, by introducing {\it ad hoc} bag function all the corrections of the energy and pressure functions of SQM have been maintained. We also consider that the quarks are non-interacting and massless in a simplified bag model. The quark pressure therefore can be defined as
\begin{equation}\label{2.8}
{p_r}={\sum_{f=u,d,s}}{p^f}-{B},
 \end{equation}
where $p^f$ is the individual pressure of the up~$\left(u\right)$, down~$\left(d\right)$ and strange~$\left(s\right)$ quark flavors and $B$ is the vacuum energy density (also well known as the `Bag' constant) which is a constant quantity within a numerical range. In the present article we consider the value of Bag constant as $B=83~MeV/{{fm}^3}$~\cite{Rahaman2014}.

Now the individual quark pressure ($p^f$) can be defined as $p^f=\frac{1}{3}{{\rho}^f}$, where ${{\rho}^f}$ is the energy density of the individual quark flavor. Hence, the energy density, $\rho$, of the de-confined quarks inside the Bag is given by
\begin{equation}
{{\rho}}={\sum_{f=u,d,s}}{{\rho}^f}+B. \label{2.9}
\end{equation}

Using Eqs.~(\ref{2.8}) and (\ref{2.9}), we have the EOS for SQM given as
\begin{equation}
{p_r}=\frac{1}{3}({{\rho}}-4B).\label{2.10}
\end{equation}

It is observed that ignoring critical aspects of the quantum particle physics in the framework of GR several authors~\cite{1,2,3,4,5,6,7,8,Deb2019a,Deb2019b} successfully have been introduced this simplified form of the MIT Bag EOS to study stellar systems made of SQM.

To have non-singular monotonically decreasing matter density inside the spherically symmetric stellar system, following Mak and Harko~\cite{Harko2002}, we assume simplified form of $\rho$ given as
\begin{equation}\label{2.11}
\rho(r)=\rho_c\left[1-\left(1-\frac{\rho_0}{\rho_c}\right)\frac{r^{2}}{R^{2}}\right],
\end{equation}
where $\rho_c$ and $\rho_0$ are two specific constants and denote the maximum and minimum values of $\rho$ at the center and on the surface, respectively.

Now following~\cite{Moraes2017} we consider $p_t$ is related to $\rho$ by a relation given as
\begin{eqnarray}
p_{t}=c_{1}\rho+c_{2},\label{2.11a}
\end{eqnarray}
where $c_1$ and $c_2$ are purely constants.

We define the mass function of the spherically symmetric stellar system as
\begin{equation}\label{2.12}
m \left( r \right) =4\,\pi\int_{0}^{r}\!{{\rho}_{eff}} \left( r \right) {r}^{2}{dr}.
\end{equation}

At this juncture we consider the Schwarzschild metric to represent the exterior spacetime of our system given as
\begin{eqnarray}\label{2.13}
 {ds}^2=\left(1-\frac{2M}{r}\right)dt^2- \frac{{dr}^2}{\left(1-\frac{2M}{r}\right)}-r^2(d\theta^2+\sin^2\theta
d\phi^2),\nonumber\\
 \end{eqnarray}
where $M$ is the total mass of the stellar system.

Now, substituting Eq.~(\ref{2.12}) into Eq.~(\ref{3.21}) we find
\begin{eqnarray}\label{2.14}
 {{\rm e}^{-\lambda \left( r \right) }}=1-{\frac {2m(r)}{r}}.
 \end{eqnarray}

In this case, the conservation equation (\ref{eqnonconserv}) in $f(T,\mathcal{T})$ gravity takes the form as follows
\begin{equation}
-p_{r}^{{\prime}}+\frac{\nu^{{\prime}}}{2} (p_r+\rho) + \frac{2}{r}(p_t -p_r)=\frac{1}{\Big(4\pi +(1/2)\varpi\Big)}\Bigg\{\frac{\varpi \rho'}{4}-\frac{\varpi p^{{\prime}}_r}{4} -\varpi p^{{\prime}}_t\Bigg\} .\label{eq40}
\end{equation}

The essential stellar structure equations required to describe static spherically symmetric sphere in $f(R, \mathcal{T})$ gravity theory are given by
\begin{eqnarray}\label{3.24}
 \frac{dm}{dr}&=&4\pi r^2\, \rho^{\textit{eff}},\,\cr \label{3.25}
\frac{dp_r}{dr}&=&\frac{1}{\left[-1+\frac{\varpi}{16\pi+2\varpi}\left(1-\frac{d\rho}{dp_r}+4\frac{dp_t}{dp}\right)\right]}\Bigg\lbrace -\left(\rho+p_r\right)\bigg[\frac{\Big\lbrace 4\pi\, r\,p_r^{\textit{eff}}+\frac{m}{r^2}\Big\rbrace }{\left(1-\frac{2m}{r}\right)}\bigg] \Bigg\rbrace \nonumber \\ +\frac{2}{r}(p_t-p_r),
\end{eqnarray}
where the metric potential $e^{\lambda}$ has the form of Schwarzschild type.

\section{Solution to the Einstein field equations in $f(\mathbb{T}, \mathcal{T})$ gravity for stellar modeling}\label{sec5}
Substituting Eq. \eqref{2.10} in Eqs. \eqref{2.11} and \eqref{2.11a}, we have the following pressures
\begin{equation}\label{3}
p_r=\frac{1}{3}\rho_c\Big[1-\Big(1-\frac{\rho_0}{\rho_c}\Big)\frac{r^{2}}{R^{2}}\Big]-\frac{4}{3}B,
\end{equation}

\begin{equation}\label{4}
p_t=c_1\rho_c\Big[1-\Big(1-\frac{\rho_0}{\rho_c}\Big)\frac{r^{2}}{R^{2}}\Big]+c_2.
\end{equation}

Again, putting Eqs. \eqref{2.10}, \eqref{3} and \eqref{4} in Eqs. \eqref{3.23}, \eqref{3.23a} and \eqref{3.23aa}, we get the following physical parameters:
\begin{equation}\label{5}
\rho^{\textit{eff}}=10\Big(1-\frac{7r^{2}}{250}\Big)+\frac{\varpi\Big[10\Big(1-\frac{7r^{2}}{250}\Big)+\frac{1}{3}\Big\{-4+10\Big(1-\frac{7r^{2}}{250}\Big)\Big\}+4\Big\{5+20\Big(1-\frac{7r^{2}}{250}\Big)\Big\}+2\Lambda\Big]}{16\pi},
\end{equation}

\begin{equation}\label{6}
p_{r}^{\textit{eff}}=\frac{1}{3}\Big\{-4+10\Big(1-\frac{7r^{2}}{250}\Big)\Big\}+\frac{\varpi\Big[10\Big(1-\frac{7r^{2}}{250}\Big)+\frac{1}{3}\Big\{-4+10\Big(1-\frac{7r^{2}}{250}\Big)\Big\}-4\Big\{5+20\Big(1-\frac{7r^{2}}{250}\Big)\Big\}-2\Lambda\Big]}{16\pi},
\end{equation}

\begin{equation}\label{7}
p_{t}^{\textit{eff}}=5+20\Big(1-\frac{7r^{2}}{250}\Big)+\frac{\varpi\Big[10\Big(1-\frac{7r^{2}}{250}\Big)+\frac{1}{3}\Big\{4-10\Big(1-\frac{7r^{2}}{250}\Big)\Big\}+2\Big\{5+20\Big(1-\frac{7r^{2}}{250}\Big)\Big\}-2\Lambda\Big]}{16\pi}.
\end{equation}

Using Eq. \eqref{5}, we integrate Eq. \eqref{3.25} and we have the mass function as
\begin{equation}\label{13}
m(r)=\frac{r^{5}(16\pi+\varpi+4c_1\varpi)(\rho_0-\rho_c)}{20R^{2}}+\frac{\rho_c r^{3}(16\pi+\varpi+4c_1\varpi)}{12}-\frac{r^{3}\varpi(2B-6c_2-3\Lambda)}{18}+d,
\end{equation}
where $d$ is integrating constant.

With the help of Eqs. \eqref{2.14}, \eqref{13}, \eqref{3.22}, \eqref{2.11}, \eqref{3} and \eqref{4}, we have the two unknown functions of $r$ of the given metric \eqref{3.13} as
\begin{equation}\label{1}
\lambda(r)=-\ln\Big(1-\frac{r^{4}(16\pi+\varpi+4c_{1}\varpi)(\rho_0 -\rho_c)}{20R^{2}}-\frac{\rho_c r^{2}(16\pi+\varpi+4c_1 \varpi)}{6}+\frac{r^{2}\varpi(2B-6c_2-3\Lambda)}{9}-\frac{2d}{r}\Big)
\end{equation}
and
\begin{equation}\label{2}
\nu(r)=\int \!\Big\{\frac{r}{1-\frac{2m(r)}{r}}\Big[-8\pi\Big\{\Big[\frac{-\varpi(\rho-p_r-2p_t)+2\Lambda}{16\pi}+\frac{\varpi p_t}{8\pi}\Big]-p_r\Big(1+\frac{\varpi}{8\pi}\Big)\Big\}+\frac{1}{r^{2}}\Big]-\frac{1}{r}\Big\} {dr}
\end{equation}

\begin{figure}
\includegraphics[height=1.35in]{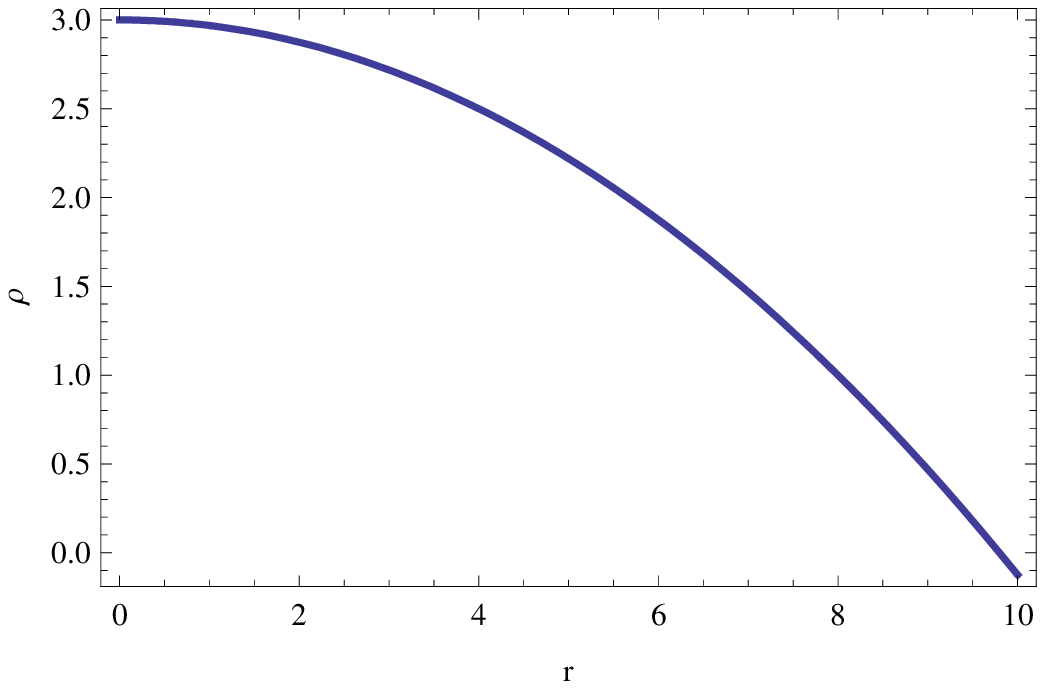}~~
\includegraphics[height=1.35in]{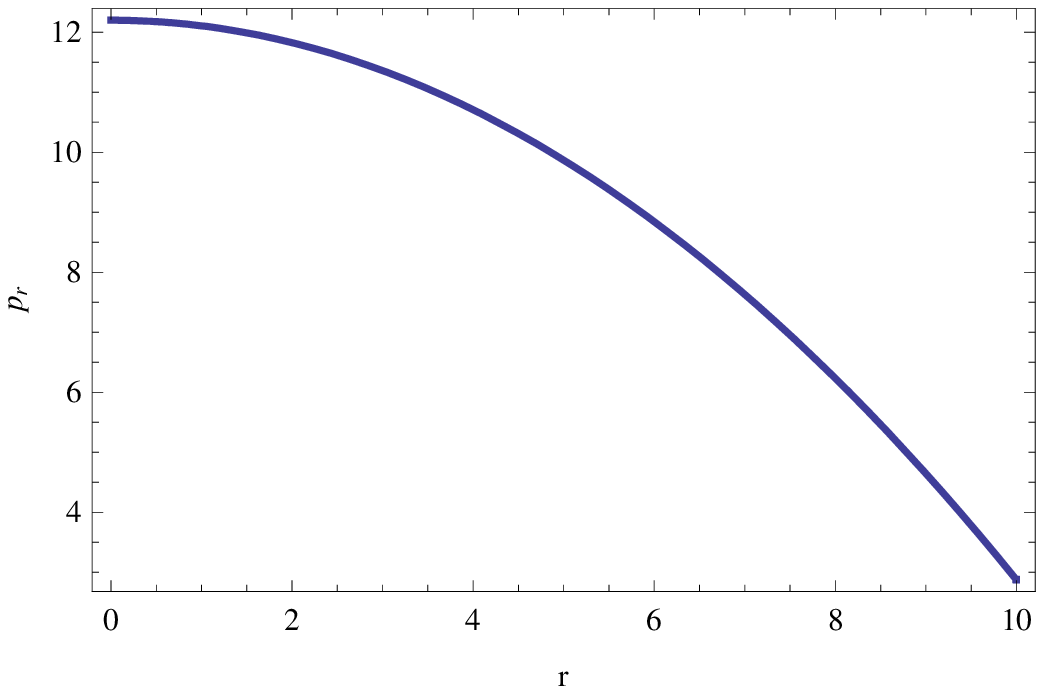}~~
\includegraphics[height=1.35in]{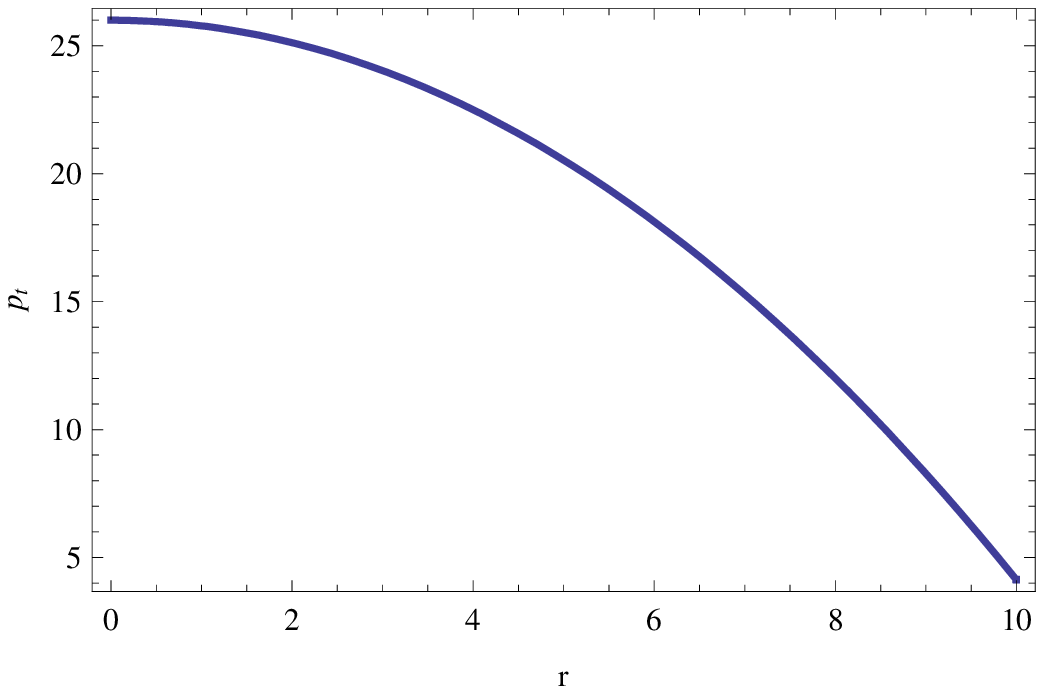}\\
\vspace{2 mm}
\textbf{Fig.~1} Variations of $\rho$, $p_r$ and $p_t$ versus $r$ (km).\\
\vspace{4 mm}
\end{figure}

\begin{figure}
\includegraphics[height=1.35in]{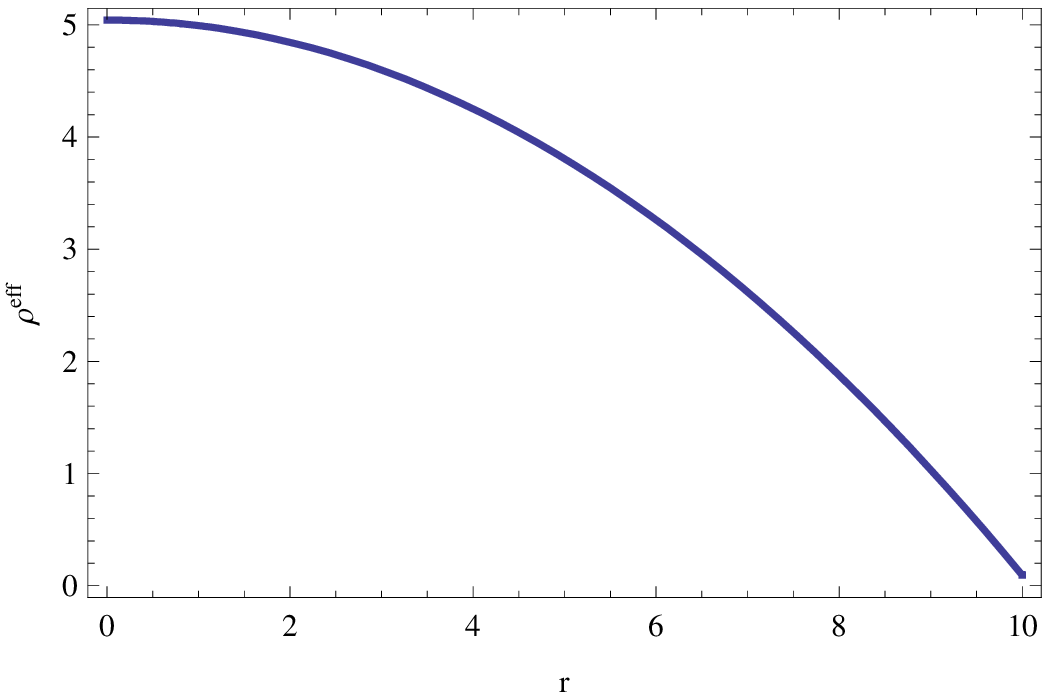}~~
\includegraphics[height=1.35in]{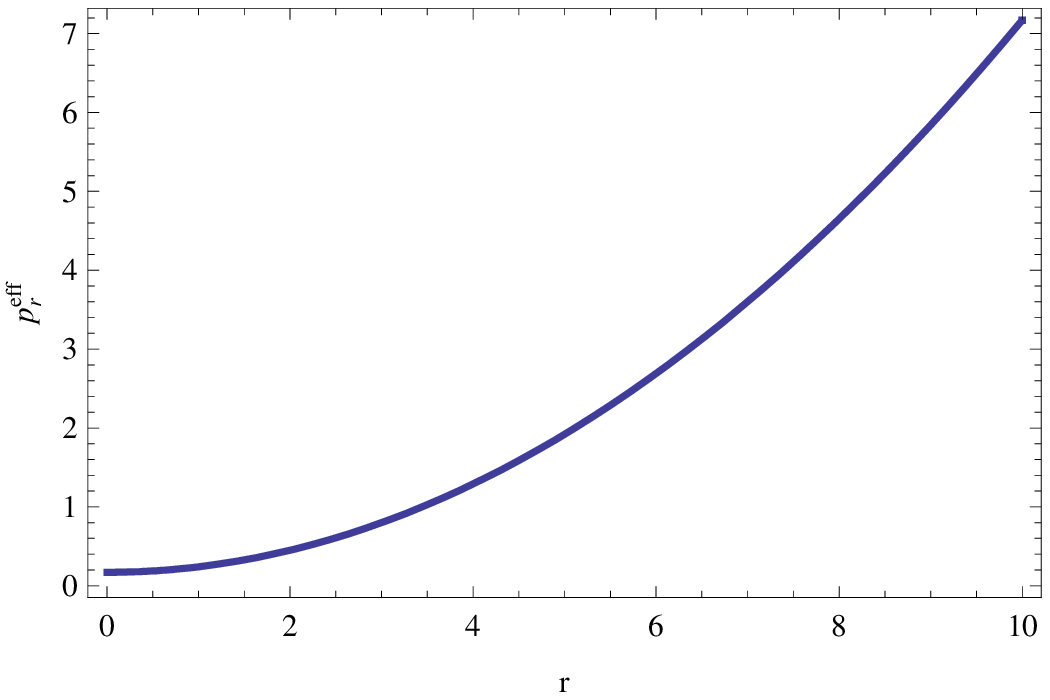}~~
\includegraphics[height=1.35in]{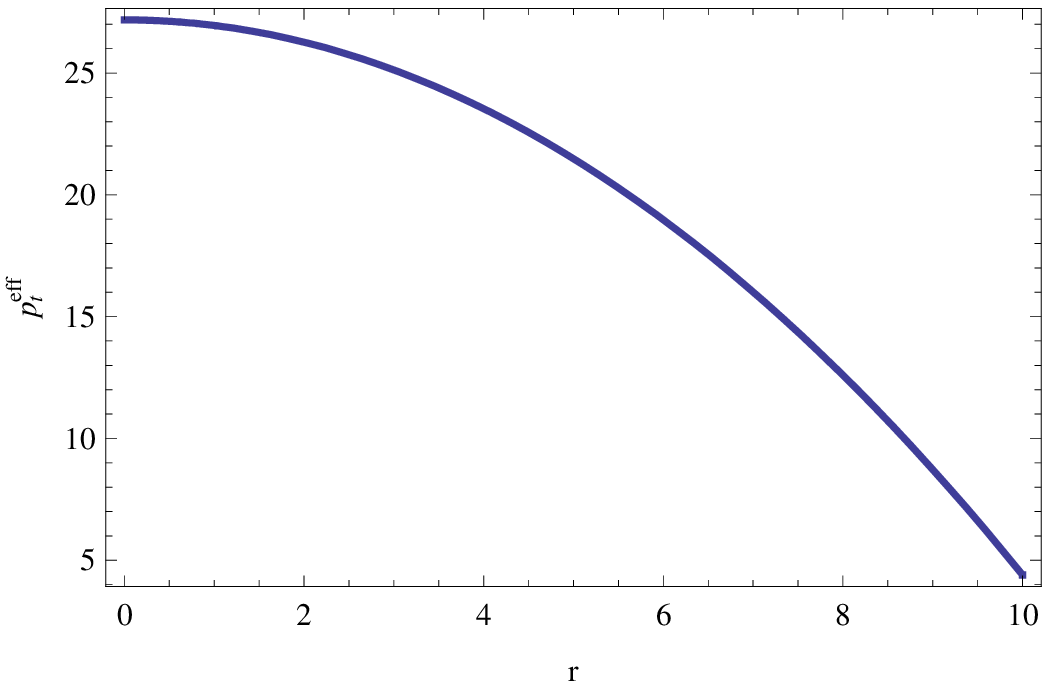}\\
\vspace{2 mm}
\textbf{Fig.~2} Variations of $\rho^{\textit{eff}}$, $p_{r}^{\textit{eff}}$ and $p_{t}^{\textit{eff}}$ versus $r$ (km).\\
\vspace{4 mm}
\end{figure}

\begin{figure}
\includegraphics[height=1.35in]{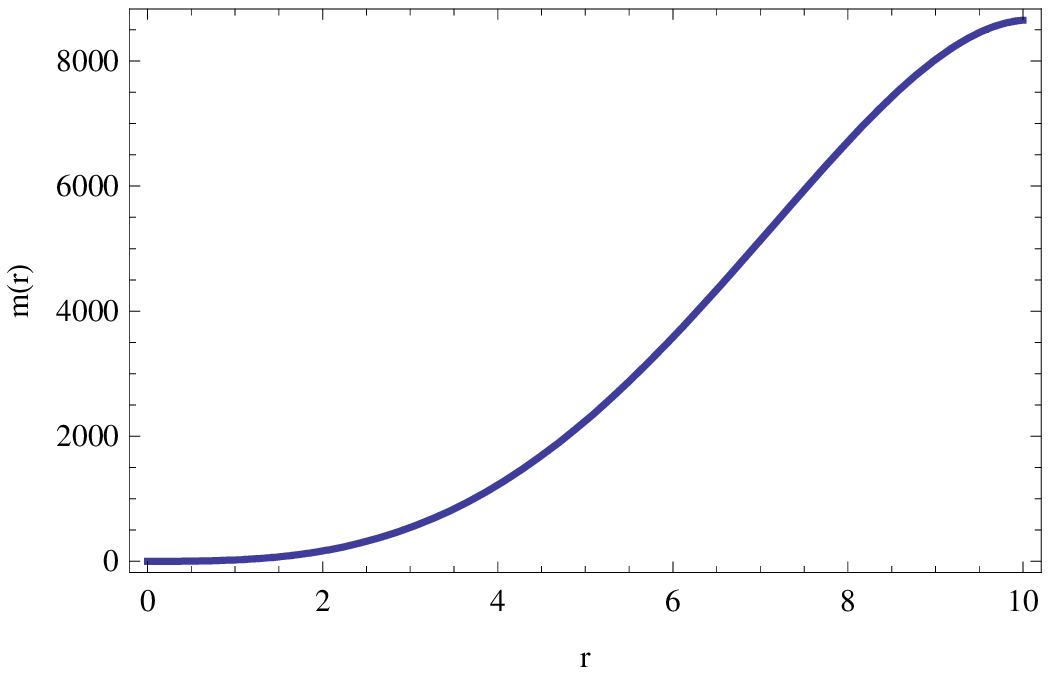}~~
\includegraphics[height=1.35in]{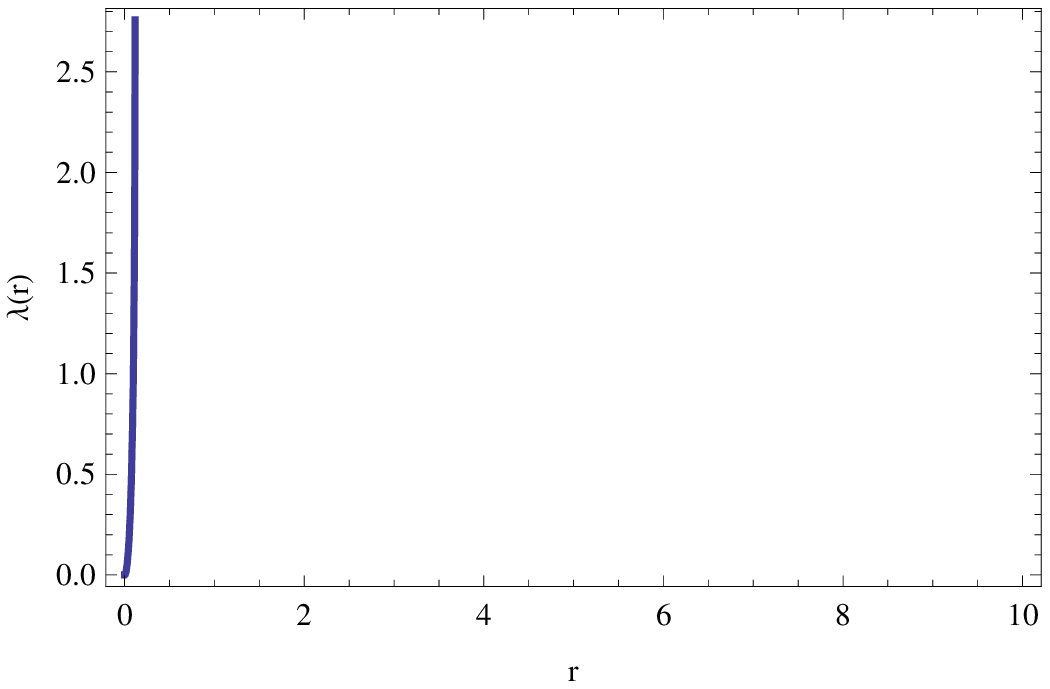}~~
\includegraphics[height=1.35in]{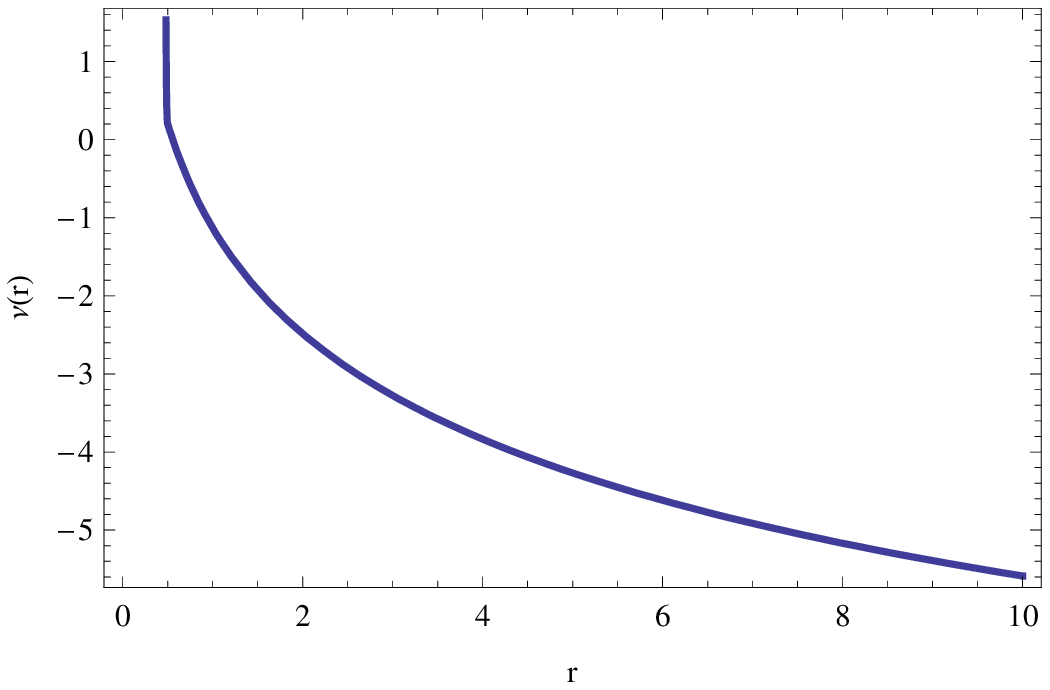}\\
\vspace{2 mm}
\textbf{Fig.~3} Variations of $m(r)$, $\lambda(r)$ and $\nu(r)$ versus $r$ (km).\\
\vspace{4 mm}
\end{figure}

We have plotted the Figs. {1} - {5} of all the physical quantities $\rho$, $p_r$, $p_t$, $\rho^{\textit{eff}}$, $p_{r}^{\textit{eff}}$, $p_{t}^{\textit{eff}}$, $m(r)$, $\lambda(r)$ and $\nu(r)$ with respect to $r$. Since $r \rightarrow 0$, $m(r) \rightarrow 0$ in Fig. {3}, so we can infer that the mass function is regular at the origin.

\section{Physical features of  strange (quark) stars in $f(\mathbb{T},\mathcal{T})$ gravity}\label{sec6}
In this Section, we study some physical features of the compact star, in order to examine the physical validity and stability of the system in the $f(\mathbb{T},\mathcal{T})$ gravity.

\subsection{Energy conditions}\label{subsec6.1}
According to relativistic classical field theories of gravitation, anisotropic fluid model must obey the energy conditions. There is often a linear relationship between the energy density and pressure of the matter by the name (i) Null Energy Condition (NEC), (ii) Weak Energy Condition (WEC), (iii) Dominant Energy Condition (DEC), (iv) Strong Energy Condition (SEC)~\cite{SKMYKGSRDD16,JPDL93} as follows:
\begin{equation}
\begin{array}{ll}
$\textbf{NEC:}$~~\rho\geq0;\\
$\textbf{WEC:}$~~\rho+p_r\geq0,~~\rho+p_t\geq0;\\
$\textbf{DEC:}$~~\rho\geq|p_t|,~~\rho\geq|p_r|;\\
$\textbf{SEC:}$~~\rho+p_r\geq0,~~\rho+p_t\geq0,~~\rho+p_r+2p_t\geq0.
\end{array}
\end{equation}

\begin{figure}
\includegraphics[height=1.35in]{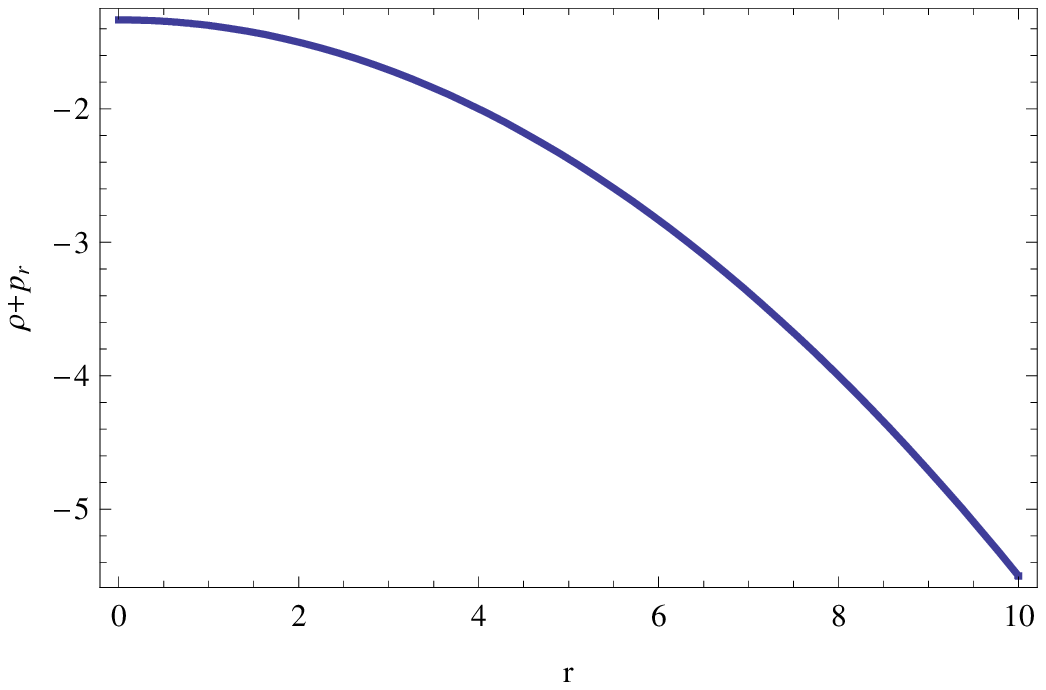}~~
\includegraphics[height=1.35in]{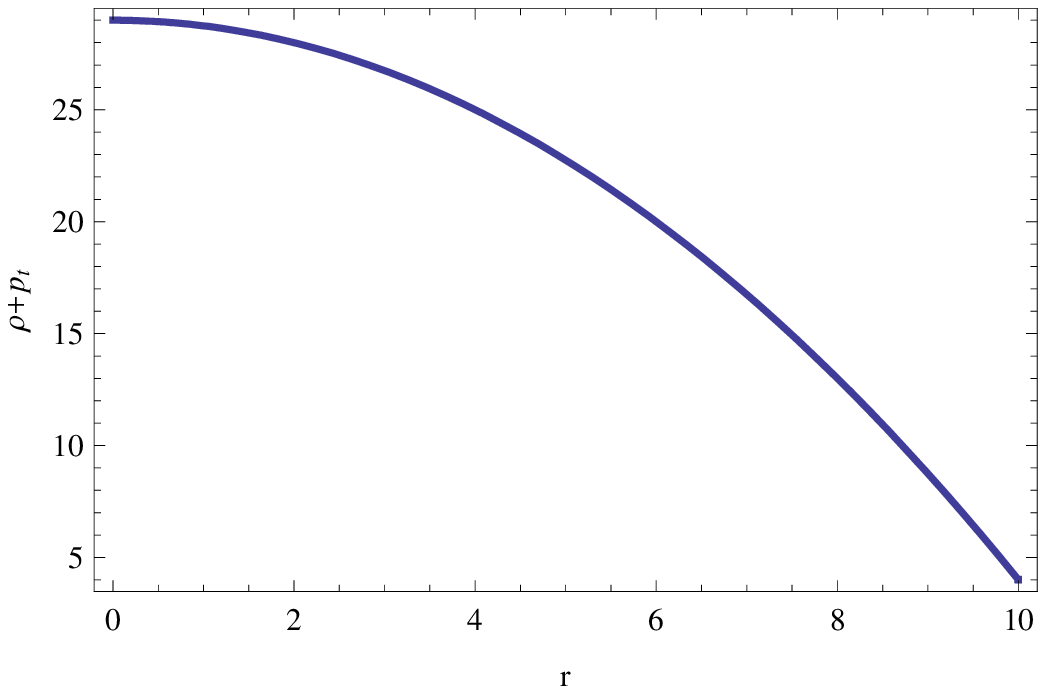}~~
\includegraphics[height=1.35in]{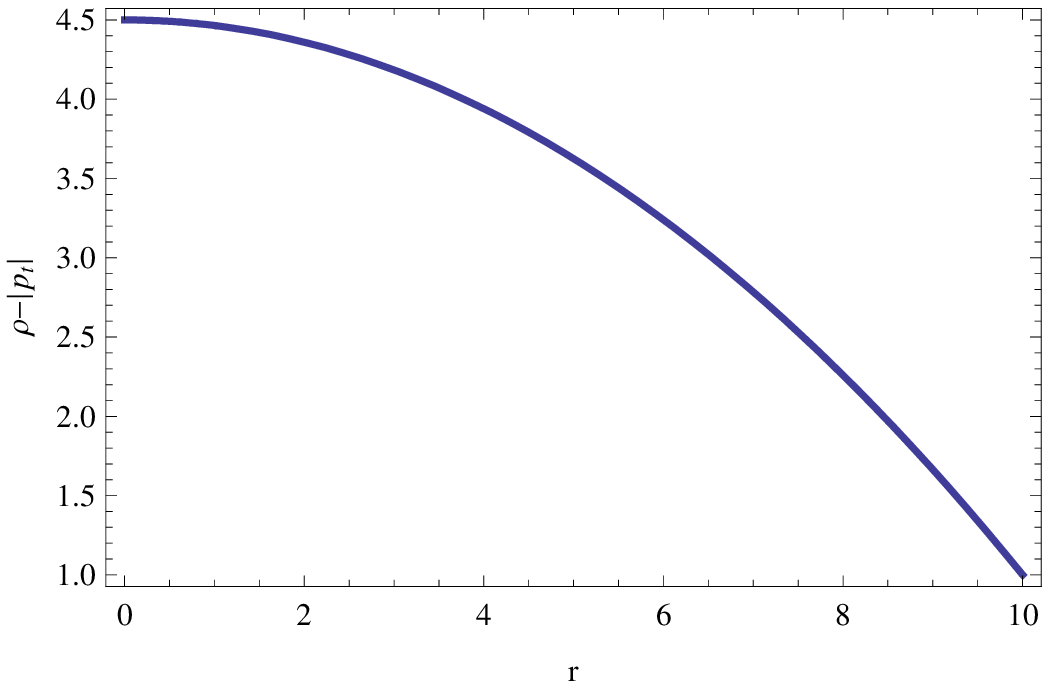}\\
\vspace{2 mm}
\textbf{Fig.~4} Variations of $\rho+p_r$, $\rho+p_t$ and $\rho-p_t$ versus $r$ (km).\\
\vspace{4 mm}
\end{figure}

\begin{figure}
\includegraphics[height=1.35in]{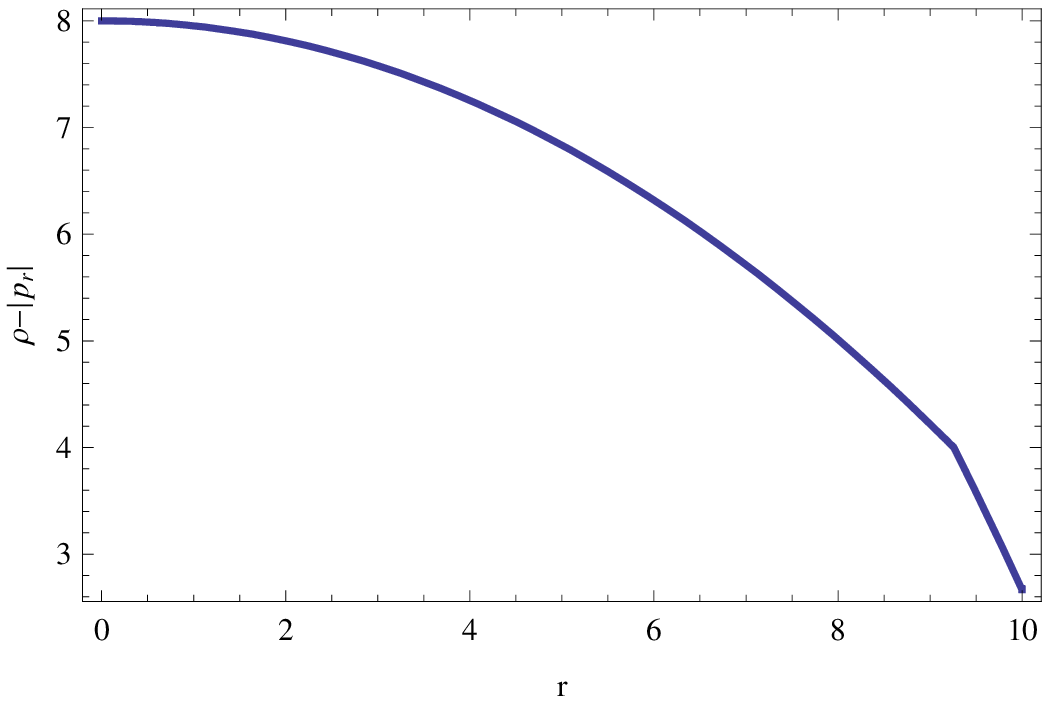}~~
\includegraphics[height=1.35in]{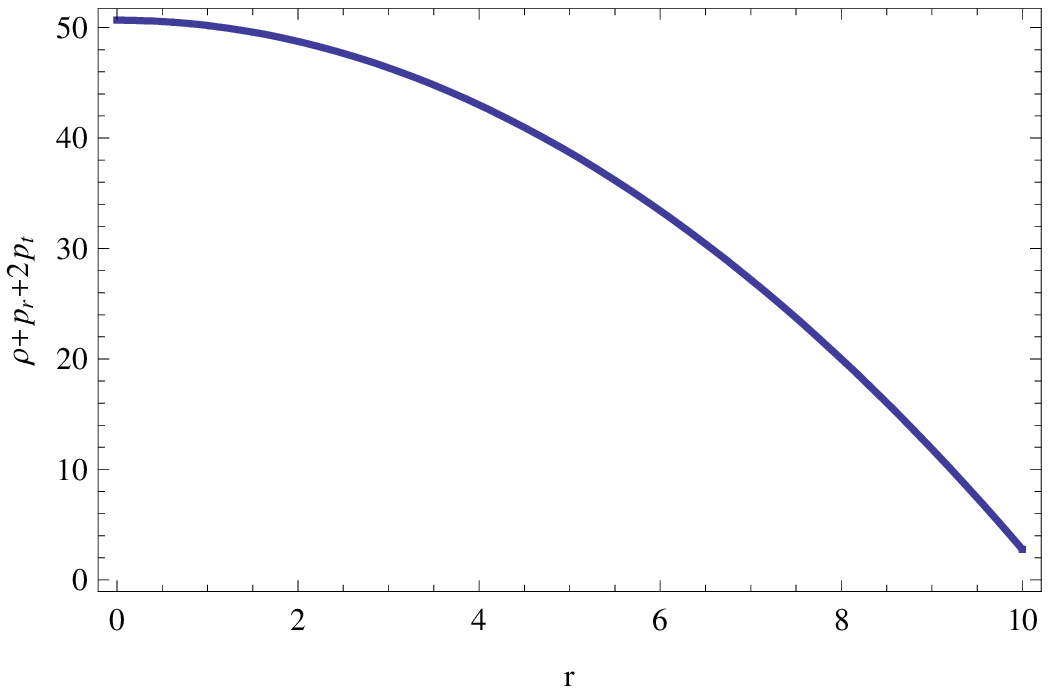}\\
\vspace{2 mm}
\textbf{Fig.~5} Variations of $\rho-p_r$ and $\rho+p_r+2p_t$ versus $r$ (km).\\
\vspace{4 mm}
\end{figure}

From Figs. {1} and {3} - {5}, we can conclude that the conditions NEC, WEC and SEC are satisfied for all values of parameters but the third one DEC is satisfied only for $c_1<1$ for our anisotropic fluid model.

\subsection{Mass-radius relation}\label{subsec6.2}
According to Bhar~\cite{1BP14}, the compactness of the anisotropic fluid model is given by the compactification factor
\begin{equation}\label{14}
\left.
\begin{array}{ll}
u(r)=\frac{m(r)}{r}\\\\
~~~~~=\frac{r^{4}(16\pi+\varpi+4c_1\varpi)(\rho_0-\rho_c)}{20R^{2}}+\frac{\rho_c r^{2}(16\pi+\varpi+4c_1\varpi)}{12}-\frac{r^{2}\varpi(2B-6c_2-3\Lambda)}{18}+\frac{d}{r}
\end{array}
\right.
\end{equation}
and the twice of the compactification factor~\cite{BHA59} is followed by
\begin{equation}\label{15}
2u(r)=\frac{r^{4}(16\pi+\varpi+4c_1\varpi)(\rho_0-\rho_c)}{10R^{2}}+\frac{\rho_c r^{2}(16\pi+\varpi+4c_1\varpi)}{6}-\frac{r^{2}\varpi(2B-6c_2-3\Lambda)}{9}+\frac{2d}{r}.
\end{equation}

\begin{figure}
\includegraphics[height=1.35in]{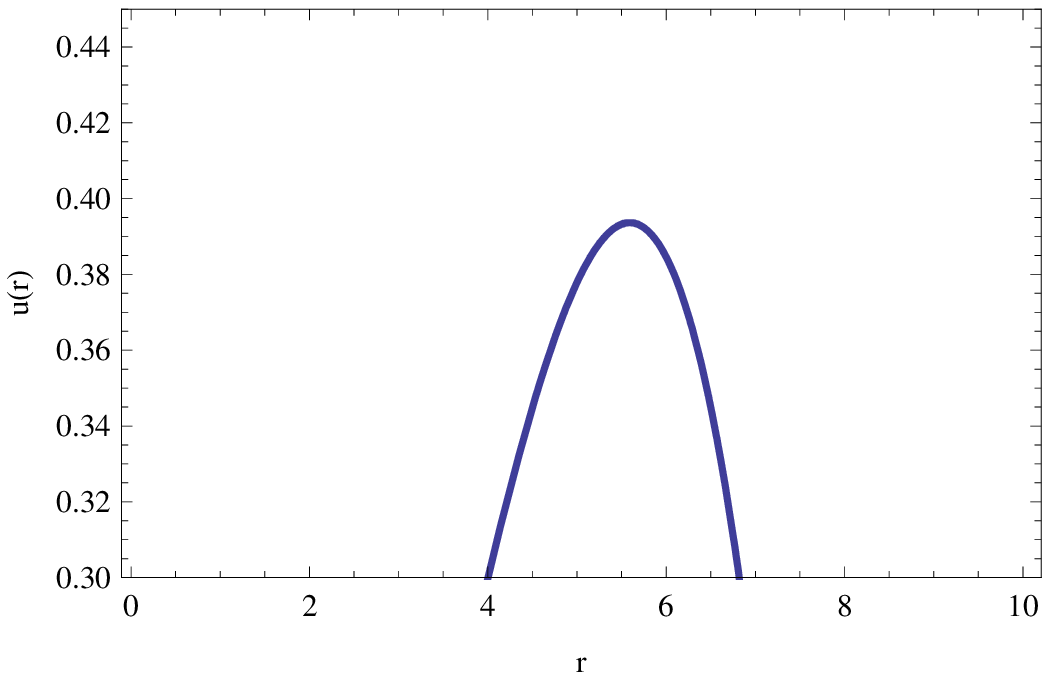}~~
\includegraphics[height=1.35in]{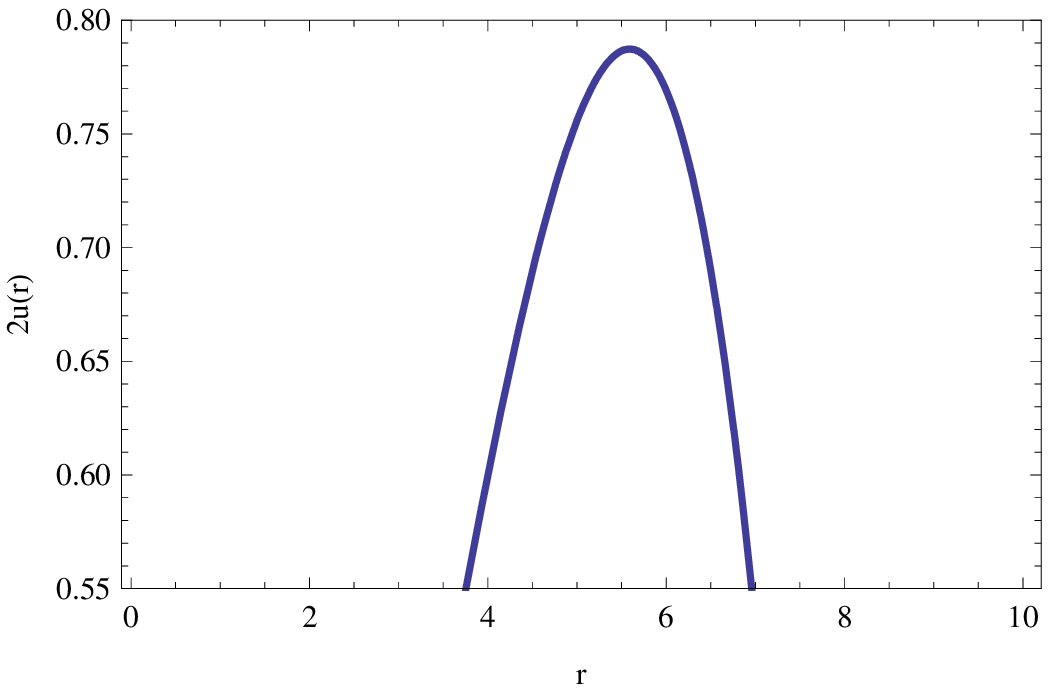}\\
\vspace{2 mm}
\textbf{Fig.~6} Variations of $u(r)$ and $2u(r)$ versus $r$ (km).\\
\vspace{4 mm}
\end{figure}

Now Fig. {6} indicates that $u(r)$ is lying between the range $\frac{1}{4}$ and $\frac{1}{2}$. We also note that the factor $2u(r)$ follows
the maximum allowed value $\frac{8}{9}$ for our proposed model.

\subsubsection{Modified TOV equation in $f(\mathbb{T}, \mathcal{T})$ gravity theory}\label{subsubsec6.3.1}
To check the hydrostatic equilibrium of our proposed model, the Tolman-Oppenheimer-Volkoff (TOV)~\cite{TRC39,OJRVGM39} condition is very useful which is followed by the Eq. \eqref{eq40}. There exist four different forces, such as the gravitational force ($F_g$), hydrostatic force ($F_h$), anisotropic force ($F_a$) and an extra force in connection to $f(\mathbb{T},\mathcal{T})$ gravity ($F_e$ ) provided in Eq. \eqref{eq40} and can be rewritten as
\begin{equation*}
F_g+F_h+F_a+F_e=0,
\end{equation*}
where
\begin{equation}\label{6.31}
\left.
\begin{array}{ll}
F_g=\frac{\nu^{\prime}}{r}(\rho+p_r), \\\\
F_h=-p_{r}^{\prime}, \\\\
F_a=\frac{2}{r}(p_t-p_r), \\\\
F_e=-\frac{1}{4\pi+\frac{\varpi}{2}}\Big(\frac{\varpi\rho^{\prime}}{4}-\frac{\varpi p_{r}^{\prime}}{4}-\varpi p_{t}^{\prime}\Big).
\end{array}
\right\}.
\end{equation}

\begin{figure}
\includegraphics[height=1.35in]{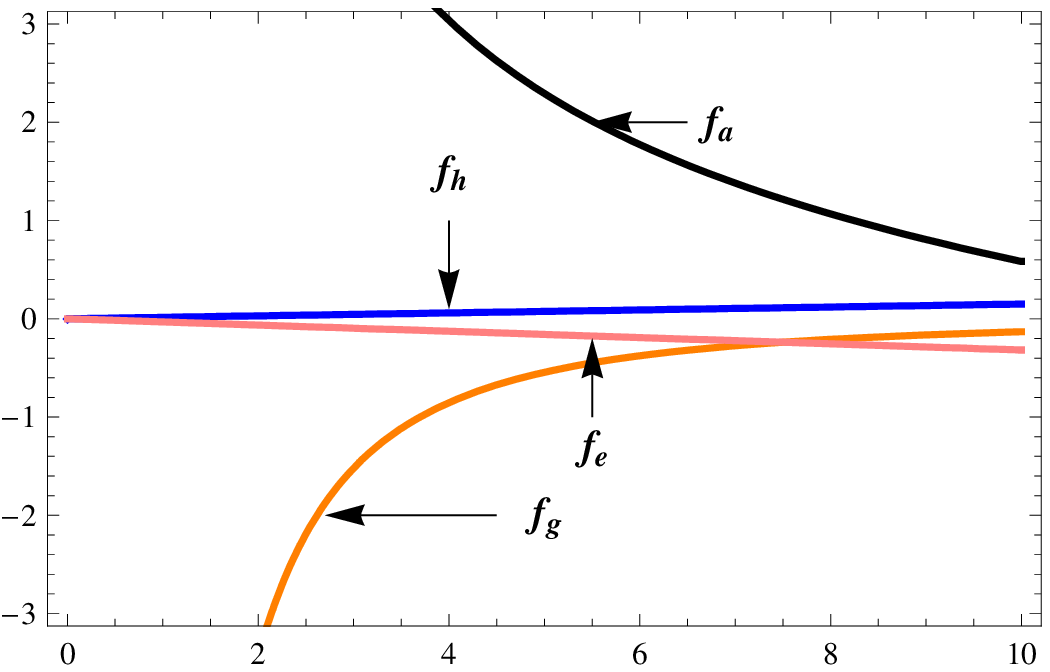}\\
\vspace{2 mm}
\textbf{Fig.~7} Variations of $F_g$, $F_h$, $F_a$ and $F_e$ versus $r$ (km).\\
\vspace{4 mm}
\end{figure}

Based on Eq. \eqref{6.31}, we have drawn Fig. {7} which indicates that our propounded anisotropic fluid model is in hydrostatic equilibrium state under the combination of considered forces.

\subsubsection{Principle of causality}~\label{6.3.2}
Anisotropic fluid stellar model will be physically admissible if we pay special attention in making of boundaries of the radial and the transversal speeds of sound in $(0,1)$ within the matter distribution, i.e., they are less than the speed of light $c$ (in relativistic geometrized units, the speed of light $c$ becomes $1$). This is known as the Causality condition which is based on the concept of Cracking condition as provided by Herrera~\cite{HL92}. Here, the radial and transversal speeds of sound of our model are
\begin{equation}\label{8}
v_{sr}^{2}=\frac{dp_r}{d\rho}=\frac{1}{3},
\end{equation}

\begin{equation}\label{9}
v_{st}^{2}=\frac{dp_t}{d\rho}=c_1.
\end{equation}

From Eqs. \eqref{8} and \eqref{9}, we observe that $0\leq v_{sr}^{2}\leq 1$ and $0\leq v_{st}^{2}\leq 1$ if $0\leq c_1 \leq 1$. So, the Causality condition is preserved inside the anisotropic strange star in $f(\mathbb{T},\mathcal{T})$.

\subsubsection{Adiabatic index}
Adiabatic index is a crucial feature to investigate the stability of an anisotropic fluid star model. It has been proposed that the radial adiabatic index should be greater than $\frac{4}{3}$ which represents the rigidity of the EOS parameter of the model~\cite{CS64,HHHW75,HWSKO76,BI96}. The radial adiabatic index is defined as
\begin{equation}
\Gamma_r=\Big(1+\frac{\rho}{p_r}\Big)\frac{dp_r}{d\rho}=\frac{4(\rho-B)}{3(\rho-4B)}.
\end{equation}

From Fig. {8}, we note that $\Gamma_r>\frac{4}{3}$ and it is monotonically increasing function at all interior points of our model and therefore, the proposed model is stable.

\subsection{Cracking condition}
There is an another way to check the stability of the system through Herrera cracking condition~\cite{HL92,HAHLA07}. Here, we check whether the square of radial speed of sound exceeds the square of transversal speed of sound or not for this anisotropic fluid star model. So, this criterion sets up
\begin{equation}\label{10}
-1\leq v_{st}^{2}-v_{sr}^{2}\leq1=
\left\{
\begin{array}{ll}
-1\leq v_{st}^{2}-v_{sr}^{2}\leq0~~~~~~$for potentially stable model$\\
~~0\leq v_{st}^{2}-v_{sr}^{2}\leq1~~~~~~$for potentially unstable model$
\end{array}
\right.
\end{equation}

Using Eq. \eqref{10}, we calculate the difference of two speeds and they are
\begin{equation}\label{11}
v_{st}^{2}-v_{sr}^{2}=c_1-\frac{1}{3},
\end{equation}

\begin{equation}\label{12}
|v_{st}^{2}-v_{sr}^{2}|=|c_1-\frac{1}{3}|.
\end{equation}

Eqs. \eqref{11} and \eqref{12} indicate that if $0\leq c_1\leq\frac{1}{3}$ then $-1\leq v_{st}^{2}-v_{sr}^{2}\leq0$ and $0\leq |v_{st}^{2}-v_{sr}^{2}|\leq1$. This means that we have potentially stable anisotropic star fluid model.

\subsection{Redshift}\label{subsec6.4}
Bohmer et al.~\cite{1BCGHT06} proposed that there exists an arbitrarily large surface redshift but must be less than $5$ for an anisotropic fluid model in the presence of the cosmological constant. The surface redshift can be defined as
\begin{equation}\label{13}
z_s=\frac{1}{\sqrt{1-\frac{2m(r)}{r}}}-1.
\end{equation}

\begin{figure}
\includegraphics[height=1.35in]{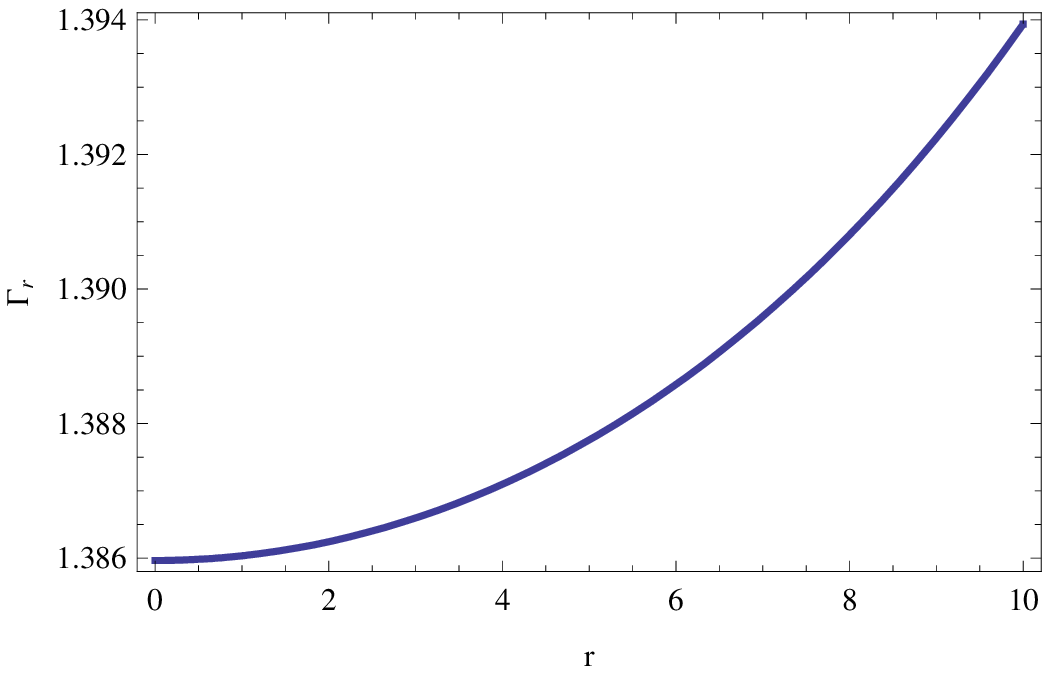}~~
\includegraphics[height=1.35in]{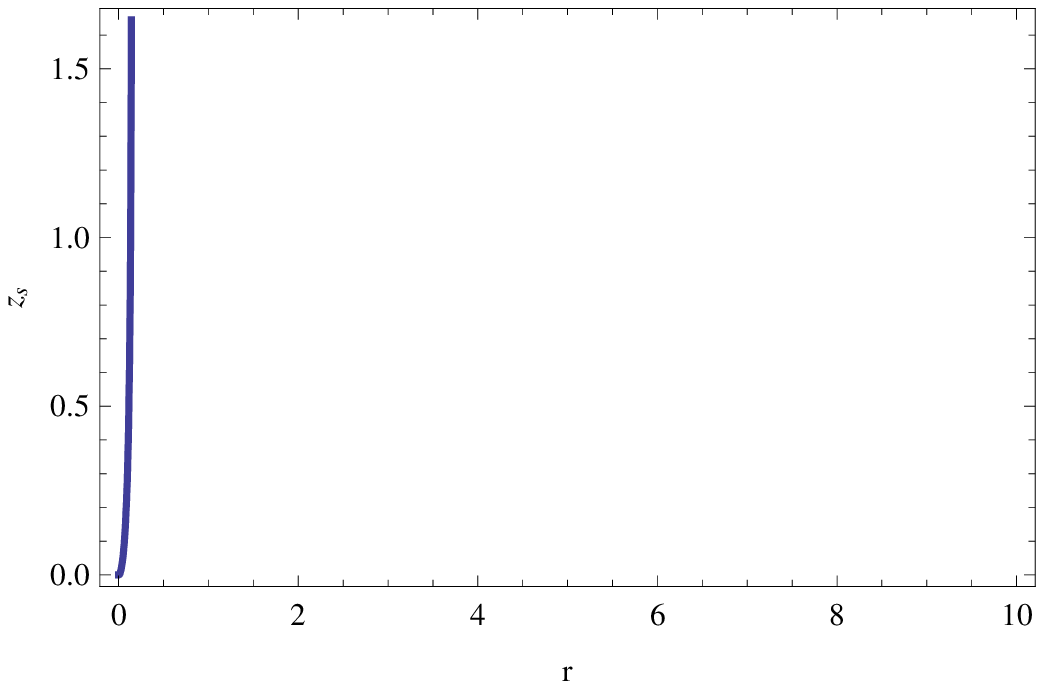}\\
\vspace{2 mm}
\textbf{Fig.~8} Variations of $\Gamma_r$ and $z_s$ versus $r$ (km).\\
\vspace{4 mm}
\end{figure}

We have plotted the figure for the surface redshift with the help of Eq. \eqref{13} in Fig. {8} which shows that the surface redshift is an increasing function starting from the core with zero value and $z_s \leq 5$ always. So, our proposed model is quite realistic.

\section{Discussions and conclusions}\label{sec7}
In this work, we present mathematical models and properties of strange star in the background of $f(\mathbb{T}, \mathcal{T})$ gravity in Einstein-Maxwell spacetime. We develop the equations of motion using anisotropic property within the spherically symmetric strange star and thereafter explore the physical features like energy conditions, mass-radius relations, modified TOV equations, principle of causality, adiabatic index, redshift and stability analysis of our model.

The salient features of the presented model can be elaborated through the figures as follows:

(1) We have plotted the Figs. {1} - {5} of all physical quantities $\rho$, $p_r$, $p_t$, $\rho^{\textit{eff}}$, $p_{r}^{\textit{eff}}$, $p_{t}^{\textit{eff}}$, $m(r)$, $\lambda(r)$ and $\nu(r)$ with respect to $r$.

(2) Since $r\rightarrow 0$, $m(r)\rightarrow 0$ in Fig. {3} so one can notice that the mass function is regular at the origin.

(3) From Figs. {1}, {3}, {4} and {5}, we can conclude that the energy conditions, viz. NEC, WEC and SEC are satisfied for all values of the physical parameters, however DEC is satisfied only for $c_1<1$ for our anisotropic fluid model.

(4) Fig. {6} indicates that $u(r)$ is lying between the range $\frac{1}{4}$ and $\frac{1}{2}$. Also, the $2u(r)$ follows the maximum allowed value $\frac{8}{9}$ for our proposed model.

(5) Fig. {7} indicates that our propounded anisotropic fluid model is in hydrostatic equilibrium state under the combination of considered forces.

(6) We note that $0\leq v_{sr}^{}\leq 1$ and $0\leq v_{st}^{2}\leq 1$ if $0\leq c_1 \leq 1$. So, the Causality condition is preserved inside the anisotropic strange star in $f(\mathbb{T},\mathcal{T})$.

(7) From Fig. {8}, we have noted that $\Gamma_r>\frac{4}{3}$ and it is monotonically increasing function at all the interior points of our model and hence provides confirmation on stability of our model.

(8) If $0\leq c_1\leq\frac{1}{3}$ then we have $-1\leq v_{st}^{2}-v_{sr}^{2}\leq0$ and $0\leq |v_{st}^{2}-v_{sr}^{2}|\leq1$. It means that we have potentially stable anisotropic star fluid model.

(9) Fig. {8} shows that the surface redshift is an increasing function starting from core zero and the condition $z_s\leq 5$ follows always.

So, based on the features as shown by the figures we can conclude that our propounded model is quite realistic and appeals to further study of theoretically observable features as well as observational signatures of its predicted compact objects. In this context we would like to point out that in the present investigation, following Rahaman et al.~\cite{Rahaman2014}, we have consider the value for Bag constant as $B=83~MeV/{{fm}^3}$. However, it seems that this value is quite arbitrary and needs specific range with lower and upper values~\cite{Aziz2019}. This means that further extension of the present work will provide more realistic stellar model to validate with the observed physical signatures.

\section*{Acknowledgments} The work by MK on astrophysical effects of modified gravty was supported by grant of the Russian Science Foundation (Project No-18-12-00213). SR and UD are thankful to the authority of Inter-University Centre for Astronomy and Astrophysics, Pune, India for providing them Visiting Associateship under which a part of this work was carried out.

\end{document}